\documentclass[12pt]{article}
\pdfoutput=1
\usepackage[utf8]{inputenc}
\usepackage{graphicx}
\usepackage{amssymb}
\usepackage{amsmath}
\usepackage{latexsym}
\usepackage[usenames]{color}
\usepackage{cite}
\usepackage{wasysym}
\definecolor{darkblue}{cmyk}{0.9,0.9,0,0}
\definecolor{darkred}{rgb}{0.6,0,0.3}
\usepackage{graphicx} 
\allowdisplaybreaks
\usepackage[setpagesize=false,pagebackref=false, linktocpage, bookmarksopen=true, colorlinks=true, linkcolor=darkblue,citecolor=darkblue,urlcolor=darkblue]{hyperref}
\usepackage{hyperref}

\renewcommand{\thefootnote}{\arabic{footnote}}
\newcommand{\s}{\sigma}
\newcommand{\epl}[1]{\epsilon_{#1}}
\newcommand{\epu}[1]{\epsilon^{#1}}
\newcommand{\brss}[2]{\langle \s_{#1} \s_{#2}\rangle}
\newcommand{\brst}[3]{\langle \s_{#1} \theta_{#2}^{#3}\rangle}
\newcommand{\yinv}[1]{(y_{#1}^{-1})}
\allowdisplaybreaks

\def\eqref#1{(\ref{#1})}

\newcommand{\beq}{\begin{equation}}
\newcommand{\eeq}{\end{equation}}
\begin{document}
\thispagestyle{empty}

\renewcommand{\thefootnote}{\fnsymbol{footnote}}
\setcounter{page}{1}
\setcounter{footnote}{0}
\setcounter{figure}{0}
%%%%%%%%%%%%%%%%%%%%%%%%%%%%%%%%%%%%%%%%%%%%%%%%%%%%%%%%%%%%%%%%%%%%%%%%%%%%%%%%%%%%%%%%%%%%%%%%%%%
\begin{center}
$$$$
{\Large\textbf{\mathversion{bold}
Non-planar data of $\mathcal N=4$ SYM}}

\vspace{1.3cm}

\textrm{Thiago Fleury$\,^{\textcolor[rgb]{0.9,0,0}{a}}$ and Raul Pereira$\,^{\textcolor[rgb]{0,0.6,0}{b}}$}
\\ \vspace{1.2cm}
\footnotesize{\textit{
$^{\textcolor[rgb]{0.8,0,0}{a}}$ 
International Institute of Physics, Federal University of Rio Grande do Norte,  \\ Campus Universit\'ario,
Lagoa Nova, Natal, RN 59078-970, Brazil
\vspace{1mm} \\
$^{\textcolor[rgb]{0,0.6,0}{b}}$ 
School of Mathematics and Hamilton Mathematics Institute, \\ Trinity College Dublin,
Dublin 2, Ireland
}  
\vspace{4mm}
}

\par\vspace{1.5cm}

\textbf{Abstract}\vspace{2mm}

\end{center}

The four-point function of length-two 
half-BPS operators in $\mathcal{N}=4$ SYM receives 
non-planar corrections starting at four loops. 
Previous work relied on the analysis of symmetries and  logarithmic divergences to fix the integrand up to four constants.
In this work, we compute those undetermined coefficients and fix the integrand completely by using
the reformulation of $\mathcal{N}=4$ SYM 
in twistor space. The final integrand 
can be written as a combination of finite conformal integrals and we 
have used the method of asymptotic expansions 
to extract non-planar anomalous dimensions 
and structure constants for twist-two operators up to spin eight.
Some of the results were already know in the literature and we 
have found agreement with them.

\noindent

\setcounter{page}{1}
\renewcommand{\thefootnote}{\arabic{footnote}}
\setcounter{footnote}{0}
\setcounter{tocdepth}{2}
\newpage
\tableofcontents

\parskip 5pt plus 1pt   \jot = 1.5ex

%%%%%%%%%%%%%%%%
\section{Introduction\label{sec:intro}}

The correlation functions of length-two 
half-BPS operators (also known as $20^{\prime}$ operators) in $\mathcal{N}=4$ SYM have been studied extensively 
in the literature both at weak \cite{WeakI,WeakII,HiddenSimmetries,Integrand,
PlanarTenLoops} and strong coupling \cite{StrongI,StrongII,StrongIII,StrongIV,StrongV,Vasco14,RastelliZhouFour,AprileI,Alday1,Alday2,Alday3,Shai1,FiveStrong,Shai2}. 
It is well known that the two- and three-point functions
of those operators are protected by supersymmetry 
\cite{StrongII,Nonrenormalization3pointsII} while higher-point functions
receive non-trivial corrections.
Each loop order of the correlation function can be computed through the Lagrangian insertion procedure, and because the $20'$ operator and the Lagrangian belong to the same supermultiplet, the integrand exhibits a hidden permutation symmetry \cite{HiddenSimmetries}. 
Imposing also conformal symmetry and restrictions from logarithmic divergences of the correlator in diverse OPE limits, the planar integrand of the four-point correlator has been fixed up to ten loops
\cite{Integrand,
PlanarTenLoops}. 
It is however important to note that knowledge of the integrated correlator is still quite incomplete since conformal integrals are generally not known
starting from four loops. 
Meanwhile, at the non-planar level much less is known.
It is easy to see that at one and 
two loops the non-planar corrections are zero since
it is not possible to draw any non-planar Feynman graph
at these loop orders. However, the vanishing of the three-loop correction
is non-trivial as it results from cancellations of different terms. Finally, at four loops 
the integrand is 
known to be a linear combination of four polynomials with constant 
coefficients. One of the main results 
of this work is the determination of those four undetermined coefficients. We expect that the methods described below can 
be adapted to the calculation of other correlators as well. 
      
One of the motivations for computing the four-loop 
non-planar integrand is that it allows to extract non-planar OPE data by taking a double coincidence limit. 
In this work we have computed both anomalous dimensions and structure constants for twist-two operators  
up to spin eight. 
Even at the planar level, where the tools of integrability are more developed, this method is the best way we currently have for obtaining structure constants of unprotected operators at a high loop order
\cite{ThreeLoopEdenAsymptotic,VascoKonishiFour,
EdenFourLoop,RaulFiveLoopKonishi}. 
In contrast, it was only recently that a direct two-loop perturbative computation of planar structure constants was performed \cite{BianchiThreeI,BianchiThreeII}. 
The double OPE limit of each conformal integral can be taken by using the method of asymptotic expansions
described for example in \cite{SmirnovRegions, ThreeLoopEdenAsymptotic}. 
With this method the integration domains are split into distinct regions, which correspond to the different scales of the problem. Effectively, the conformal four-point integrals can be rewritten in terms of two-point integrals, which are much more tractable. Note that the pseudo-conformal integrals arising in high-loop integrands can also be approximated with this method, as long as one does not assume dependence on cross-ratios. This is however not relevant to this work, as we can write the four-point function in terms of convergent integrals only.

The non-planar anomalous dimensions of 
twist-two operators up to spin six 
were also computed in the series
of papers \cite{FourLoopNonPlanarKonishi,VelizhaninI,
Velizhaninspin6} by a direct Feynman diagrammatic 
calculation and we have found agreement 
with them\footnote{In fact there is a mismatch for the spin six data. Our coefficient for $\zeta_3$ is ten times the
one in \cite{Velizhaninspin6}. Because the mismatch is
simple, we believe that there is a typo on the paper \cite{Velizhaninspin6}.}. It would be very interesting to find a closed expression
for the non-planar anomalous dimensions of twist-two
operators for any spin. We expect the result to be given in terms of harmonic sums and Riemann zeta values and to obey the principle of uniform transcendentality. Unfortunately the data obtained in this work 
is not enough to fix the expression for general spin
even if one restricts to a smaller basis consisting only
of binomial harmonic sums \cite{BinomialHarmonicSums}. 
The knowledge of the non-planar anomalous dimension for generic spin would allow to compute the non-planar cusp anomalous
dimension analytically and also take the BFKL 
limit. These results are important for understanding 
non-planar integrability or possible formulations
of a non-planar quantum spectral curve \cite{QuantumEspectralCurve}, see 
\cite{ShotaGiombiNonPlanar} for progress in this direction.
One should also stress that the 
non-planar cusp anomalous dimension 
was computed numerically   
by studying Sudakov form-factors with a suitable rewriting in terms of uniformly transcendental integrals \cite{NumericalNonplanarI,
	NumericalNonplanarII, NumericalNonplanarIII}.

Another motivation for computing non-planar 
structure constants is to further the understanding of 
non-planar integrability.  Three-point functions can be computed at the planar level
as a product of two integrable hexagon form-factors \cite{BKV}, and it was later understood  that higher-point functions can also be decomposed into a weighted product of hexagon form-factors, both at the planar and non-planar level \cite{Hexagonalization, EdenTessellating, HandlingI, HandlingII,
EdenNonPlanar, OctagonsNonPlanar}.
The integrability setup was tested at two loops
for long operators and at one loop for short operators
such as the $20^{\prime}$.
In the case of four-point functions of length-two 
operators,  
all the non-planar corrections in the integrability
setup come from the stractification procedure described in \cite{HandlingII}. More specifically, one embeds the tree-level planar graphs in higher 
genus surfaces and properly subtracts possible 
boundary terms. It would be interesting to test the stratification
procedure at higher loops (at the moment four loops 
seems a difficult task) or at least understand 
why non-trivial non-planar corrections 
to this correlator first show up at four loops.

This paper is organized as follows. In the remaining part
of the introduction, we review what is known 
in the literature about four-point functions of $20^{\prime}$ operators up to four loops. 
In section  \ref{sec:Twistors} we introduce the reformulation of $\mathcal{N}=4$ SYM in twistor space
and explain how to compute correlation functions in that framework. We then describe the strategy used to fix the four-loop non-planar integrand.
The OPE analysis of the correlator is performed in section
\ref{sec:OPE}, where we present the non-planar OPE data of twist-two low-spin operators. 
We conclude in section \ref{sec:conclusion} and refer the reader to the appendices for conventions and examples
of analytic computations using twistors.

\subsection{Four-point function of $20^{\prime}$  operators}\label{se:ReviewKoretall} 

We will now review some well-known results about correlation functions in $\mathcal{N}=4$ SYM. We will focus our attention on the four-point function of length-two half-BPS operators, which can be written as
\begin{equation}\label{20'}
\mathcal{O}_{\mathbf{20'}}(x_i,y_i)=y_i  \,  y_j  \, {\rm{Tr}} (  \Phi^i  \Phi^ j) (x_i) \,,
\end{equation}
with $\Phi^i$ the six 
real scalars of the theory and $y_i$ a null polarization vector which projects the operator into the symmetric traceless representation. The correlator admits a double expansion
in the effective coupling constant $a = g^2_{\mathrm{YM}} N_c/(4 \pi^2)$ and in the number of colours 
$N_c$    
\begin{equation}
G_4 = \langle \mathcal{O}_{\mathbf{20'}}(x_1,y_1) \mathcal{O}_{\mathbf{20'}}(x_2,y_2)
 \mathcal{O}_{\mathbf{20'}}(x_3,y_3) \mathcal{O}_{\mathbf{20'}}(x_4,y_4) \rangle= 
\sum_{\ell \geq 0}   \sum_{g \geq 0} 
\frac{a^{\ell}}{N_c^{2g}}  \,  G_4^{(g,\ell)}
(x_i,y_i) \, .  
 \end{equation} 
Starting at one loop, $\mathcal{N}=4$ superconformal symmetry 
\cite{EdenPartialNonRenormalization} 
implies that at any order in $N_c$ the correlator $G_4^{(g,\ell)}$ factorizes in the following way
\begin{equation}
G_4^{(g,\ell>0)}(x_i,y_i) = 2(N_c^2 -1)\times R(x_i,y_i)  \times
F^{(g, \ell)} (x_i)   \, , 
\label{eq:factorizationproperty} 
\end{equation}
where all depence on the polarization vectors of the external operators is encoded in the $R$ factor 
\begin{align}
R&(x_i,y_i) =
\frac{y_{12}^4 y_{34}^4}{x_{12}^2 x_{34}^2} + 
\frac{y_{13}^4 y_{24}^4}{x_{13}^2 x_{24}^2} +
\frac{y_{14}^4 y_{23}^4}{x_{14}^2 x_{23}^2} +  
\frac{y_{12}^2 y_{23}^2 y_{34}^2 y_{14}^2}{x_{12}^2 x_{23}^2 x_{34}^2 x_{14}^2}
(x^2_{13} x^2_{24} - x^2_{12} x^2_{34} - x^2_{14} x^2_{23}) \nonumber\\
& +  \frac{y_{12}^2 y_{13}^2 y_{24}^2 y_{34}^2}{x_{12}^2 x_{13}^2 x_{24}^2 x_{34}^2} (x^2_{14} x^2_{23}
-x^2_{12} x^2_{34} - x^2_{13} x^2_{24}) +
\frac{y_{13}^2 y_{14}^2 y_{23}^2 y_{24}^2}{x_{13}^2 x_{14}^2 x_{23}^2 x_{24}^2} (x^2_{12} x^2_{34} -
x^2_{14} x^2_{23} - x^2_{13} x^2_{24})  \, .  
\label{eq:DefinitionR1234}  
\end{align}
Notice that all dynamical information is contained in the functions $F^{(g,\ell)}$, which crucially multiply all six $R$-symmetry structures. We can therefore work with a particular choice of polarizations where only $y_{12}$ and $y_{34}$ are non-zero, so that a single term in the $R$ factor survives. This will significantly reduce the number of graphs to be computed, as explained in more detail later in the next section, but one can  unambiguously reconstruct the correlator for generic polarizations.

Loop corrections for the correlator can be obtained by the Lagrangian insertion procedure, where the integrand of the $\ell$-loop four-point function is viewed as a Born-level $(4+\ell)$-point function. We can then rewrite the dynamical function as 
\begin{equation}
F^{(g, \ell)} (x_i)
= \frac{x_{12}^2 x_{13}^2 x_{14}^2 x^2_{23} x^2_{24}
x^2_{34}}{\ell!}  \int 
d^4 x_5 \ldots d^4 x_{4+\ell} \, f^{(g,\ell)} 
(x_1, \ldots, x_{4+\ell}) \, ,  
\label{eq:TheFormofF}
\end{equation}
where the integrand $f^{(g,\ell)}$ carries conformal weight +4 in all external and internal points. Furthermore, an analysis of the possible OPE singularities indicates that the integrand is a rational function which diverges at most as a simple pole in the coincidence limit $x_{ij}\rightarrow 0$, which implies that we can rewrite it as 
\begin{equation}
 f^{(g,\ell)} (x_1, \ldots, x_{4+\ell}) = \frac{P^{(g,\ell)} (x_1, \ldots, x_{4+\ell})}{\prod_{1 \leq i < j \leq 4+\ell} 
 x^2_{ij}} \, . 
\label{eq:deff}
\end{equation}
Finally, $P^{(g,l)}$  is a linear combination of polynomials which have the following properties:
\begin{enumerate}
\item It is homogeneous in $x^2_{ij}$. 

\item It has conformal weight $-(\ell-1)$ at each point.  \label{property2}

\item It is invariant under the permutation of all its arguments, i.e. under the group $S_{4+\ell}$.   \label{property3}

\end{enumerate} 
Property \ref{property2} follows from conformal symmetry, while property  \ref{property3} reflects a hidden permutation symmetry, which follows from the fact that the Lagrangian operator is in the same supermultiplet of the external $ \mathbf{20'}$ operators. 

At each loop order there is a finite number of polynomials $P_i^{(\ell)}$ that satisfy the properties listed above and  each $P^{(g,l)}$  is a linear combination of those  with constant coefficients.  Note that the properties above are independent of $g$, so the basis $P_i^{(\ell)}$ which solves these constraints can be used to construct the numerator of the integrand at any order of the genus expansion.
At one-, two-, three- and four-loops there are 1, 1, 4 and 32 independent polynomials respectively. We will not write them explicitly in this work so we refer the reader to reference
 \cite{Integrand}.   In order to fix the integrand completely, one only needs to find 
the coefficient multiplying
each polynomial. A powerful method to fix these coefficients is to study the asymptotic behaviour of the  correlator either in the   
Euclidean double short-distance limit, where both $x_1 \rightarrow x_2$ and $x_{3} \rightarrow x_4$, or 
the Minkowski  light-cone 
limit where $x^2_{12}, x^2_{23}, x^2_{34}, x^2_{41} \rightarrow 0$. In these limits the logarithm of the correlator must develop soft logarithmic singularities, which imposes strong constraints on the coefficients.  These constraints, together with the conformal Gram determinant relations\footnote{The Gram 
determinant condition expresses the fact that in $d$ dimensions there are only $d$ independent 
vector positions $x^{\mu}$. This constraint can be imposed in a conformally invariant way, leading 
to a conformal Gram determinant, see Appendix B of 
\cite{Integrand} for details.} were  
powerful enough to fix the planar four-loop result and to
reduce the non-planar corrections at four loops to only four unknown
coefficients. 

Before writing down the form of the 
non-planar integrand, let us clarify the classification 
of the polynomials regarding their planarity. 
For each homogeneous polynomial $P^{(\ell)}_i$ obeying the conformal and permutation symmetries described above, it is possible to associate a graph $f^{(\ell)}_i$ via  \eqref{eq:deff}. Each graph has $4+\ell$ vertices and they are connected by propagators, which are the elements of the denominator  left in \eqref{eq:deff} after cancellation with factors from $P^{(\ell)}_i$, while the remaining numerator factors plays no role in the $f$-graph. A nice feature of these $f$-graphs is that they are  
in a sense related to the usual Feynman graphs and we 
can naturally associate a genus to them in the standard
way. 
More specifically, it was argued in \cite{Integrand}
that, apart from the singular one-loop case, the
$1/N_c^{2g}$ correction to the correlator $G_4$
is given by $f$-graphs whose genus is at most $g$. 
In conclusion, each polynomial 
$P^{(\ell>1)}_i$ is assigned a genus equal to that of its associated $f$-graph and it can only contribute to the integrand $P^{(g,\ell)}$ if the genus obtained does not exceed $g$.

The analysis of the non-planar integrand in \cite{Integrand} showed that 
corrections to $G_4$ first appear at four loops, but the constraints were not sufficient to fix it uniquely. At genus one, the integrand is given up to four undetermined coefficients  \begin{equation}
P^{(1,4)}(x_1, \ldots, x_8) = c_1 Q_{1}(x_i) + 
c_2 Q_{2}(x_i) +c_3 Q_{3}(x_i) +c_4 Q_{4}(x_i) \, .
\end{equation}
Each  term is a linear combination of the 32 four-loop polynomials $P^{(4)}_i$ (see equations (5.9) and (C.1) of \cite{Integrand} for definitions)
\begin{equation}\label{Qs}
Q_k(x_1, \ldots, x_8) = \sum_{j=1}^{32} q_{k,j} P_{j}^{(4)}(x_1, \ldots, x_8) \, , 
\end{equation}
with
\begin{align}
q_1 &= \{ 0^{26},1,0^{5}\} \, , \nonumber \\ 
 q_2 &= \{0,-2,2,-8,0,-6,0^4,2,-2,1,0^4,-2,-1,1,6,0^{11}\} \, , \label{eq:TheDefiningPolynomials} \\
 q_3 &= \{0,-2,-3,4,1,-6,-2,2,-4,0,2,-2,1,1,-1,2,1,0^{15} \} \, , \nonumber \\
q_4 &= \{-8, -14,10,-8,8, -18,0,3,-4,-4,0,-2,3,4,-2,6,0,-2,-4,4,0,12,2,-2,2,0^7\} \, .
\nonumber 
\end{align}
 where the short-hand notation $0^n$ corresponds to a list of $n$ zeros. 

One of the results of this work is the determination 
of the coefficients $c_i$.  We obtained $c_1=c_2=c_3=0$ and 
$c_4=-6$, and in that way we fixed the non-planar integrand at four loops completely. The method we have used 
relies on the reformulation of $\mathcal{N}=4$ SYM
in twistor space, which is the subject of the next section.      

\section{Twistors}\label{sec:Twistors} 

In this section, we first review how to compute 
correlation functions of the stress-tensor supermultiplet  in 
$\mathcal{N}=4$ SYM using twistor space, see 
\cite{Twistors} for further details.
One of the advantages of this formalism is that 
each Feynman diagram in twistor space has manifest $\mathcal{N}=4$ superconformal 
symmetry apart from some reference 
twistor. Then we 
explain how the four-loop non-planar calculation
was performed for a particular polarization of the external operators.  
The necessary graphs were generated with 
the open source program Sage \cite{Sage}. 
   
\subsection{$\mathcal{N}$=4 SYM in twistor space}
\label{sec:N4TwistorSpace} 

The supertwistor space \cite{PenroseTwistor,FerberSuperTwistors} is the complex 
projective superspace $\mathbb{CP}^{3|4}$.
An element  $\mathcal{Z}^A$  of this space has four bosonic  
and four fermionic coordinates and it is defined 
up to the equivalence relation 
$\mathcal{Z}^A \sim c \mathcal{Z}^A$, with
$c \in \mathbb{C}^{*}$. These variables are parametrized in the following way 
\begin{equation}\label{supertwistor}
\mathcal{Z}^A =( \lambda_{\alpha}, \mu^{\dot{\alpha}}, \chi^a) \, ,  
\end{equation}
with $\alpha, \dot{\alpha}=1,2$ and $\chi^a$, $a=1, \ldots,4$, the fermionic coordinates. A nice property of these variables
is that they transform linearly under the action of all generators of the complexified super
conformal group $SL(4|4; \mathbb{C})$, see 
for example
\cite{BerkovitsTwistorTransform,BullimoreBook}  
 for the explicit form of the generators.
In addition, these variables can be related to 
the usual superspace variables  
$(x^{\alpha \dot{\alpha}}, \theta^{a \alpha}, \bar{\theta}_{\dot{a}}^{\dot{\alpha}})$. We are interested here only in the chiral superspace, 
i. e.  we are going to set  all
$\bar{\theta}_{\dot{a}}^{\dot{\alpha}}$ to zero, and in this case  
\begin{equation}
\mu^{\dot{\alpha}} = i  x^{\dot{\alpha} \beta} \lambda_{\beta}  \, , \quad \chi^{a} = \theta^{a \alpha} \lambda_{\alpha} \, . 
\label{eq:incidencerelations} 
\end{equation}  
These are called incidence relations 
and they map a point in chiral superspace to a 
line in supertwistor space.

The first relation in (\ref{eq:incidencerelations}) can be
understood as follows. For simplicity, let us consider the bosonic components of the supertwistors,  in which case the 
complexified conformal group is $SL(4; \mathbb{C})$.   
The supersymmetric case is a simple generalization.
The twistors $Z^I=( \lambda_{\alpha}, \mu^{\dot{\alpha}})$, with $I=1, \ldots, 4$, transform in
the fundamental representation of this bosonic group. We can define a null antisymmetric tensor $X^{I J}$ as 
(see the Appendix \ref{sec:Conventions} for conventions)
\begin{equation}
X^{IJ} = 
\left(
\begin{tabular}{cc}
$\epsilon_{\alpha \beta}$  & $- i x^{\dot{\beta}}_{\alpha}$  \\
$ i x^{\dot{\alpha}}_{\beta} $ & $- x^2 \epsilon^{\dot{\alpha}
\dot{\beta}} $ 
\end{tabular}
\right)\, , \quad X_{IJ} = \frac{1}{2} 
\epsilon_{IJKL} X^{KL} \, , \quad X_{IJ} X^{IJ}=0 \, .
\label{eq:ConditionsonX} 
\end{equation}
These tensors are also homogeneous, with $X^{IJ} \sim c X^{IJ}$, and the set of
null rays is in correspondence with the original $4d$ spacetime
coordinates $x^{\alpha \dot{\beta}}$.
This identification is known as the embedding formalism. Because of the null condition given in
(\ref{eq:ConditionsonX}), the matrix $X^{IJ}$ has rank 
two and it can be written in terms of two twistors as
\begin{equation}
X^{IJ} = Z_1^I Z_2^J - Z_1^J Z_2^I \, .
\end{equation}              
As mentioned before, this implies that a spacetime point is mapped to a line in twistor space. The line
connects the two twistors $\{Z_1^I, Z_2^J\}$ 
which have linearly independent values for 
$\lambda_{\alpha}$ and the component $\mu^{\dot{\alpha}}$ given by
by (\ref{eq:incidencerelations}).

It is possible to reformulate $\mathcal{N}=4$ SYM in
supertwistor space and in that way we gain an alternative method
for computing correlation functions of the stress-tensor multiplet. The fields of $\mathcal{N}=4$ SYM sit inside a one-form superfield $\mathcal{A}$ living in supertwistor 
space. Accordingly, the action can 
be written as a function of this superfield in the following way (see \cite{WittenTwistor,N4Twistor} for details)  
\begin{equation}
S_{\mathcal{N}=4}  =
\int_{\mathbb{CP}^{3|4}} 
\mathcal{D}^{3|4} \mathcal{Z} \wedge {\rm{Tr}} \left(
\frac{1}{2} \mathcal{A} \bar{\partial} \mathcal{A} 
- \frac{1}{3} \mathcal{A}^3   
\right)  + g^2_{\mathrm{YM}}   \int d^4 x \, d^8 \theta \;L_{int}(x,\theta) \,.
\label{eq:twistoraction}
\end{equation}
In order to perform calculations, it is convenient to choose a gauge in which a component of the superfield $\mathcal A$ vanishes in the direction of 
a reference twistor  $\mathcal{Z}_{\diamond}$, so that the kinetic term becomes quadratic and the interaction term simplifies to
\begin{equation}\label{Lint}
L_{int}(x,\theta) = - \sum_{k\geq 2} \frac{1}{k}\int D\sigma_1 \ldots D\sigma_k  \frac{\mathrm{Tr}[\mathcal A(\mathcal Z(\sigma_1)) \ldots \mathcal A(\mathcal Z(\sigma_k))]}{\langle \sigma_1 \sigma_2\rangle \ldots \langle \sigma_k \sigma_1\rangle}\,,
\end{equation}
where the superfields are integrated along a line 
in twistor space
\begin{equation}
\mathcal{Z}(\sigma_i) = \mathcal{Z}_{\alpha} \sigma_i^{\alpha} \, ,
\end{equation}
with $ \mathcal{Z}_{\alpha} $ satisfying the incidence relations (\ref{eq:incidencerelations}) for
different $\lambda_{\alpha}$.    
The measure is $D\sigma=\langle \sigma \,d\sigma\rangle$
and the bracket notation stands for $\langle \sigma_i \sigma_j \rangle=\epsilon_{\alpha \beta} \,\sigma_i^\alpha \sigma_j^\beta$. While non-trivial, it has been shown that any physical 
quantity is independent of the reference twistor $\mathcal{Z}_{\diamond}$.

The spacetime equations of motion can be obtained from
the action above by expanding the superfield $\mathcal{A}$ in components
\begin{equation}
\mathcal{A} (\mathcal{Z}, \bar{\mathcal{Z}}) = a + \chi^a \psi_a   
+ \frac{1}{2} \chi^a \chi^b \phi_{ab}
+ \frac{1}{3!} \epsilon_{abcd} \chi^{a} \chi^{b} \chi^{c}
\psi^{\prime d} + \frac{1}{4!} \epsilon_{abcd} 
\chi^{a} \chi^{b} \chi^{c} \chi^{d} a^{\prime}  \, ,
\end{equation}        
In the formula above the fields on the right-hand side depend on the bosonic twistors $Z, \bar{Z}$ parametrizing a line. Moreover $\{a, a^{\prime}\}$ are the
two gluon helicity states, $\{\psi_a, \psi^{\prime d}\}$ are the gluinos and $\phi_{ab}$ are the six scalars.   
It is very important to notice that the 
action above is chiral and it contains the 
topological term $ i F \tilde{F}$, where $F$ is the field
strength. While this term is not important in perturbation 
theory because it is a total derivative, it is going to contribute to the integrand we want to compute by introducing a term proportional to the spacetime tensor
$\epsilon^{\mu \nu \rho \lambda}$. These terms have the wrong parity and they integrate to zero \cite{EpsilonTerms}.               
 
\subsection{Correlation functions of the stress-tensor multiplet}\label{sub:RulesTwistors}

Our aim is to compute the correlation function of four $\mathbf{20'}$ operators defined in \eqref{20'}. This operator is the lowest component of the
stress-tensor supermultiplet $\mathcal{T}$, whose top component  is the  Lagrangian. The fact that this is a short multiplet implies that $\mathcal{T}$ depend only in half of the odd variables 
$\theta^{a \alpha}, \bar{\theta}_{a}^{\dot{\alpha}}$\,.        

In order to define the relevant fermionic degrees of freedom, it is convenient  
to introduce the auxiliary harmonic variables
$u^b_a \equiv (u^{+ \mathfrak{b} }_{a} , 
u^{ \mathfrak{-b^{\prime}} }_{a} )$ which 
parametrize the coset 
$SU(4)/(SU(2) \times SU(2)^{\prime} \times U(1))$.
The indices $a,b$ are fundamental indices of $SU(4)$, while $\mathfrak{b}, \mathfrak{b}^{\prime}$ are fundamental
 indices of  $SU(2)$ and $SU(2)^\prime$ respectively, with the signs indicating the $U(1)$ charge. These variables and their complex conjugates satisfy several unitary and completeness conditions which follow because $u_a^b$ is
in $SU(4)$. The harmonic variables allow us to write manifestly $SU(4)$ invariant expressions.
If one defines
\begin{equation}
\theta^{+ \mathfrak{b}}_{\alpha} = \theta^{a}_{\alpha} \,  u_{a}^{+ \mathfrak{b}} \, , \quad   \quad 
\theta^{- \mathfrak{b^{\prime} }}_{\alpha} = \theta^{a}_{\alpha} \,  u_{a}^{- \mathfrak{b^{\prime}}} \, , 
\label{eq:thetaplusminus} 
\end{equation}
then one can decompose every $\theta^a_{\alpha}$ as
\begin{equation}
\theta^a_{\alpha} = \theta^{+ \mathfrak{b}}_{\alpha} 
\bar{u}_{+ \mathfrak{b}}^a  
+\theta^{- \mathfrak{b^{\prime} }}_{\alpha} 
\bar{u}_{- \mathfrak{b^{\prime}}}^a \, , 
\end{equation}        
where $\bar{u}$ are the complex conjugates of the harmonic variables $u$. The stress-tensor superfield $\mathcal T(x,\theta^+, \bar\theta_-,u)$ depends on half of the odd variables, both chiral and anti-chiral, but it is useful to focus on the chiral part of the multiplet, where all $\bar\theta_-$ vanish, so that we have 
\begin{equation}
\mathcal{T}(x, \theta^+,u) = 
\mathcal{O}^{++++}(x)  + \ldots + (\theta^+)^4 \mathcal{L}(x) \, ,
\label{eq:stresstensorexpansion}
\end{equation}
and we have omitted the other powers of $\theta^+$ in the expansion. The field $\mathcal{L}(x)$ is the chiral Lagrangian and  $\mathcal{O}^{++++}(x) = {\rm{Tr}}(\phi^{++} \phi^{++})$  is a representation of the half-BPS operator defined in 
\eqref{20'} with $\phi^{++} = \phi^{ab} u_a^{+ \mathfrak{b}} u_b^{+ \mathfrak{c}} \epsilon_{\mathfrak{b}\mathfrak{c}}$. The connexion between
$u^a_b$ and $y_i$ is made with the following 
particular parametrisation of the harmonic variables
\begin{equation}
u^{+ a}_b = (\delta^{\mathfrak{a}}_{\mathfrak{b}},
y^{\mathfrak{a}}_{\mathfrak{b^{\prime}}} )  \, , \quad
u^{- \mathfrak{a^{\prime}}}_b = (0, \delta^{\mathfrak{a^{\prime}}}_{\mathfrak{b^{\prime}}}) 
\, , \quad \bar{u}^b_{+ \mathfrak{a}}= (\delta_a^b,0) 
\, , \quad  \bar{u}^b_{- \mathfrak{a^{\prime}}} =
( - y^{\mathfrak{b}}_{\mathfrak{a^{\prime}}}, \delta^{\mathfrak{b^{\prime}}}_{\mathfrak{a^{\prime}}}) \, ,
\label{eq:DefinitionyTwoIndices} 
\end{equation} 
with $y^2 =- y_{\mathfrak{a^{\prime}}}^{\mathfrak{b}} y^{\mathfrak{a^{\prime}}}_{\mathfrak{b}}/2$ and 
the indices $\mathfrak{a}, \mathfrak{a}^{\prime}$
are raised and lowered as usual with the epsilon tensors.    

The correlation functions $\mathcal G_n = \langle \mathcal{T}(1) \ldots  \mathcal{T}(n) \rangle$ have a series 
expansion in $\theta_i^+$ as a consequence of 
(\ref{eq:stresstensorexpansion}), 
and we can extract   
the correlation function of four $\mathbf{20'}$
operators by computing $\mathcal G_4$ and reading its lowest component, or
equivalently, by sending all $\theta_i^+$ to zero.
For four- and higher-point functions there is a dependence on the coupling $a$, which can be made precise through the Lagrangian insertion method
\begin{equation}
\frac{\partial}{\partial g^2_{\mathrm{YM}} } \mathcal G_n = \int d^4 x_{n+1} 
\,\langle \mathcal{T}(1) \ldots \mathcal T(n) \mathcal{L}(x_{n+1}) \rangle = \int d^4 x_{n+1} \,d^4 \theta^+_{n+1} \,\mathcal G_{n+1} \,,   \label{Linsertion}
\end{equation}
or, more generally, 
\begin{equation}
\frac{1}{m!} 
\frac{\partial^m \mathcal G_n}{\partial g^{2m}_{\mathrm{YM}} }  = 
 \int \prod_{i=1}^m  d^4 x_{n+i}\, d^4  \theta^+_{n+i} \, \mathcal G_{n+m} \, .   
\end{equation} 
On the other hand, from equation \eqref{eq:twistoraction} we can also derive the following insertion formula
\begin{equation}
\frac{\partial}{\partial g^2_{\mathrm{YM}} } \mathcal G_n = \int d^4 x_{n+1} \,d^8 \theta_{n+1} 
\,\langle \mathcal{T}(1) \ldots \mathcal T(n) L_{int}(x_{n+1},\theta_{n+1}) \rangle\,,
\end{equation}
which hints at the following representation of the stress-tensor superfield in twistor space
\begin{equation}
\mathcal T(x,\theta^+) = \int d^4 \theta^-\, L_{int}(x, \theta)\,.
\end{equation}
In order to extract the $\ell$-loop four-point function, we will then have to compute the following $(4+\ell)$-point correlator in twistor space
\begin{equation}\label{GfromLint}
\mathcal G_{4+\ell} = \int d^4\theta_1^- \ldots d^4 \theta_{4+\ell}^- \, \langle L_{int}(x_1,\theta_1) \ldots L_{int}(x_{4+\ell},\theta_{4+\ell})\rangle \,.
\end{equation}
Looking at equation \eqref{eq:twistoraction} we see that each $L_{int}$ comes with at least two superfield insertions along the twistor line. Moreover, according to equation \eqref{Linsertion}, we also need to integrate the $\theta^+$ variables for each of the Lagrangian insertions, which means that at $\ell$ loops we are looking for the component of $\mathcal G_{4+\ell}$ with Grassmann degree $4\ell$. Since each vertex reduces the degree by four units while each propagator increases it by four, then at $\ell$ loops we need to construct  all diagrams with $4+\ell$ vertices and $4+2\ell$ propagators. All such graphs can be generated 
by the open source Sage \cite{Sage}.

The computation of $\mathcal G_n$ at tree level is made by
summing all relevant diagrams and for each graph in twistor space we have to use the Feynman rules which were derived in \cite{Twistors}:
\begin{enumerate}

\item A propagator connecting vertices $i$ and $j$ provides a factor $d_{ij} = y_{ij}^2/x^2_{ij}$ and a colour delta function $\delta^{a_i a_j}$,
\item A bivalent vertex contributes 
a colour factor ${\rm{Tr}}( T^{a_1} T^{a_2}) = 
\delta^{a_1 a_2}$,
\item Higher-valence vertices are associated with 
the factor $R^i_{j_1\ldots j_m} {\rm{Tr}}( T^{a_1} \ldots T^{a_m}) $,
\end{enumerate}
with $T^{a}$ the generators of the gauge group and the $R$ factor defined by   
\begin{equation}
R^i_{j_1 j_2 \cdots j_k} = - \int d^4 \theta_i^- \frac{\delta^2(\langle \sigma_{i j_1} \theta_{i}^- \rangle  + A_{i j_1})\delta^2(\langle \sigma_{i j_2} \theta_{i}^- \rangle  + A_{i j_2}) \ldots \delta^2(\langle \sigma_{i j_k} \theta_{i}^- \rangle  + A_{i j_k})}{\langle 
\sigma_{i j_1} \sigma_{i j_2} \rangle \langle 
\sigma_{i j_2} \sigma_{i j_3} \rangle \ldots \langle 
\sigma_{i j_k} \sigma_{i j_1} \rangle} \, .
\label{eq:defRij1jk}
\end{equation}
The delta functions are fermionic and therefore, by construction, $R^i_{j_1 j_2 \cdots j_k}$ has Grassmann degree $2k-4$. 
The $\sigma_{ij}$ originate from the integrations in equation \eqref{Lint}, which are localised by the twistor propagators and become  
\begin{equation}
\sigma^{\alpha}_{ij} = \epsilon^{\alpha \beta} 
\frac{\langle Z_{i, \beta} Z_{\diamond} Z_{j,1} Z_{j,2} \rangle}{\langle Z_{i,1} Z_{i,2} Z_{j,1} Z_{j,2} \rangle} 
 \, , 
 \label{eq:sigmadefinition}
\end{equation}
where $Z_{j,1}$ and $Z_{j,2}$ are the bosonic components of the
twistors parametrizing a line which corresponds to 
the spacetime point $x^{\mu}_j$. Finally, by setting the fermionic components of the auxiliary supertwistor to zero we have       
\begin{equation}
A^{\mathfrak{a^{\prime}}}_{ij} = [ \langle \sigma_{ji} \theta_j^{+\mathfrak{b}} \rangle +
\langle \sigma_{ij} \theta_i^{+\mathfrak{b}} \rangle ](y^{-1}_{ij})^{\mathfrak{a^{\prime}}}_{\mathfrak{b}}  \, .  
\label{eq:DefinitionofA} 
\end{equation}

The $R$ factors defined in (\ref{eq:defRij1jk}) have several important properties and satisfy 
some identities which can be found in \cite{Twistors}. In what follows we will only need two of these 
identities. First, since the numerator of  (\ref{eq:defRij1jk}) does not depend on the ordering of $\{j_1,\ldots, j_k\}$ then the effect of a permutation $\rho$ on the indices is simply
\begin{equation}
R^i_{j_{\rho(1)}\ldots j_{\rho(k)}} = R^i_{j_1 \ldots j_k} \frac{\brss{i j_1}{i j_2} \ldots \brss{i j_k}{i j_1}}{\brss{i j_{\rho(1)}}{i j_{\rho(2)}} \ldots \brss{i j_{\rho(k)}}{i j_{\rho(1)}}} \, .
\label{eq:PermutationofTheIndices} 
\end{equation}
In this way we can rewrite $R$ factors in a canonical way and reduce the number of fermionic integrations we need to perform.
Second, a multi-index $R^i_{j_1 j_2 \cdots j_k} $ can always be factorized as follows                 
\begin{equation}
R^i_{j_1 j_2 \cdots j_k} 
= R^{i}_{j_1 j_2 j_3} R^{i}_{j_1 j_3 j_4} 
\ldots R^{i}_{j_1 j_{k-1} j_k} \, ,   
\label{eq:DecomposingR} 
\end{equation}
which implies that each diagram can be rewritten in terms of a fundamental building block $R^i_{j_1 j_2 j_3}$, which takes the following form after the integration over the $\theta^{-}_i$
\begin{equation}
R^i_{123} = - \frac{ \delta^2 \left( \langle \sigma_{i1} \sigma_{i2} \rangle A_{i3} + \langle \sigma_{i2} \sigma_{i3} \rangle A_{i1} +\langle \sigma_{i3} \sigma_{i1} \rangle A_{i2}  \right)}{\langle 
\sigma_{i 1} \sigma_{i 2} \rangle \langle 
\sigma_{i 2} \sigma_{i 3} \rangle  \langle 
\sigma_{i 3} \sigma_{i 1} \rangle} \, .   
\label{eq:FundamentalR} 
\end{equation}

Once we sum all relevant diagrams, we obtain the component of $\mathcal G_{4+\ell}$ with fermionic degree $4\ell$. And since we want to perform the $\int \mathrm d^4\theta^+$ integrations at the Lagrangian insertions to obtain the loop-level four-point function, effectively we need to send the $\theta_i^+$ at the external points to zero. 
This implies that the $A_{ij}^{\mathfrak{a^{\prime}}}$ defined in (\ref{eq:DefinitionofA})  can only give a non-zero contribution  if at least one of the indices corresponds to an internal point, which in turn means that we only have dependence on $y_{ij}^{-1}$ if at least one of the indices is from an integrated point. 

Despite the obvious simplicity of this statement, it does imply that if there is a propagator between external points $k$ and $l$, the resulting $y_{kl}^2$  factor can never be cancelled as an effect of the fermionic integrations. Consequently, if we choose a particular polarization where $y_{kl}$ vanishes, then we can neglect all diagrams which contain a propagator between those two points.
This is a great simplification because the factorized form of the correlator in equation \eqref{eq:factorizationproperty} allows us to select external polarizations such that only $y_{12}$ and $y_{34}$ are different from zero, thus greatly reducing the number of twistor space diagrams we need to evaluate. In the next subsection, we explain how to
compute a four-point function with this 
assumption. It is possible to perform several intermediate analytical
computations as well and we give examples in the Appendix \ref{sec:analytical}.

\subsection{Four-loop four-point function}

\begin{figure}[t]
\centering
\includegraphics[width=0.6\textwidth]{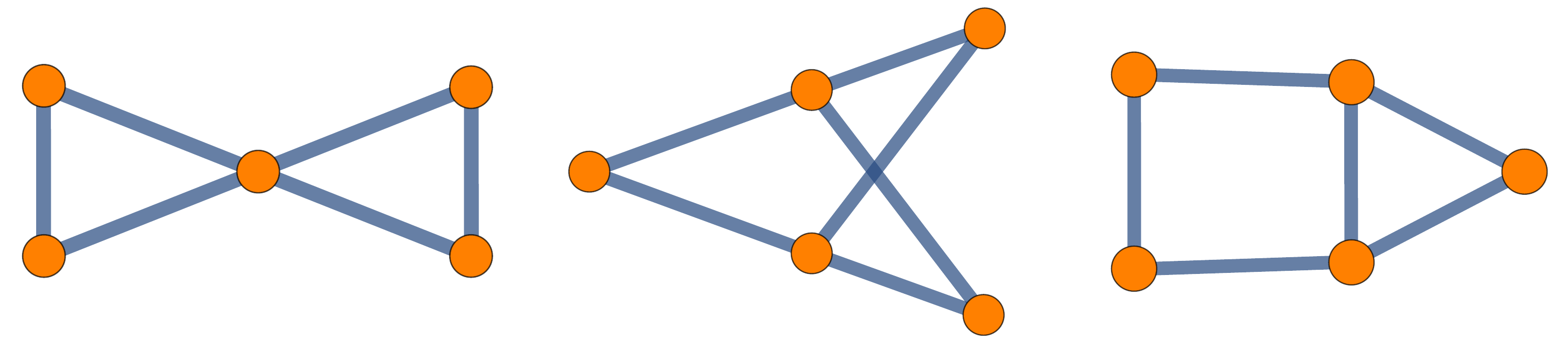}
\caption{The skeleton graphs used for the one-loop computation of a four-point function. 
The ribbon graphs are obtained
from these by adding colours traces.  Only the first graph contributes to the particular choice of polarization,
and the only configuration which does not vanish is when the middle vertex corresponds to the integrated Lagrangian insertion. One can easily compute it analytically with equation  (\ref{eq:analiticalBotie}).}
\label{fig:OneLoopGraphs} 
\end{figure}

\begin{figure}[t]
\centering
\includegraphics[width=0.7\textwidth]{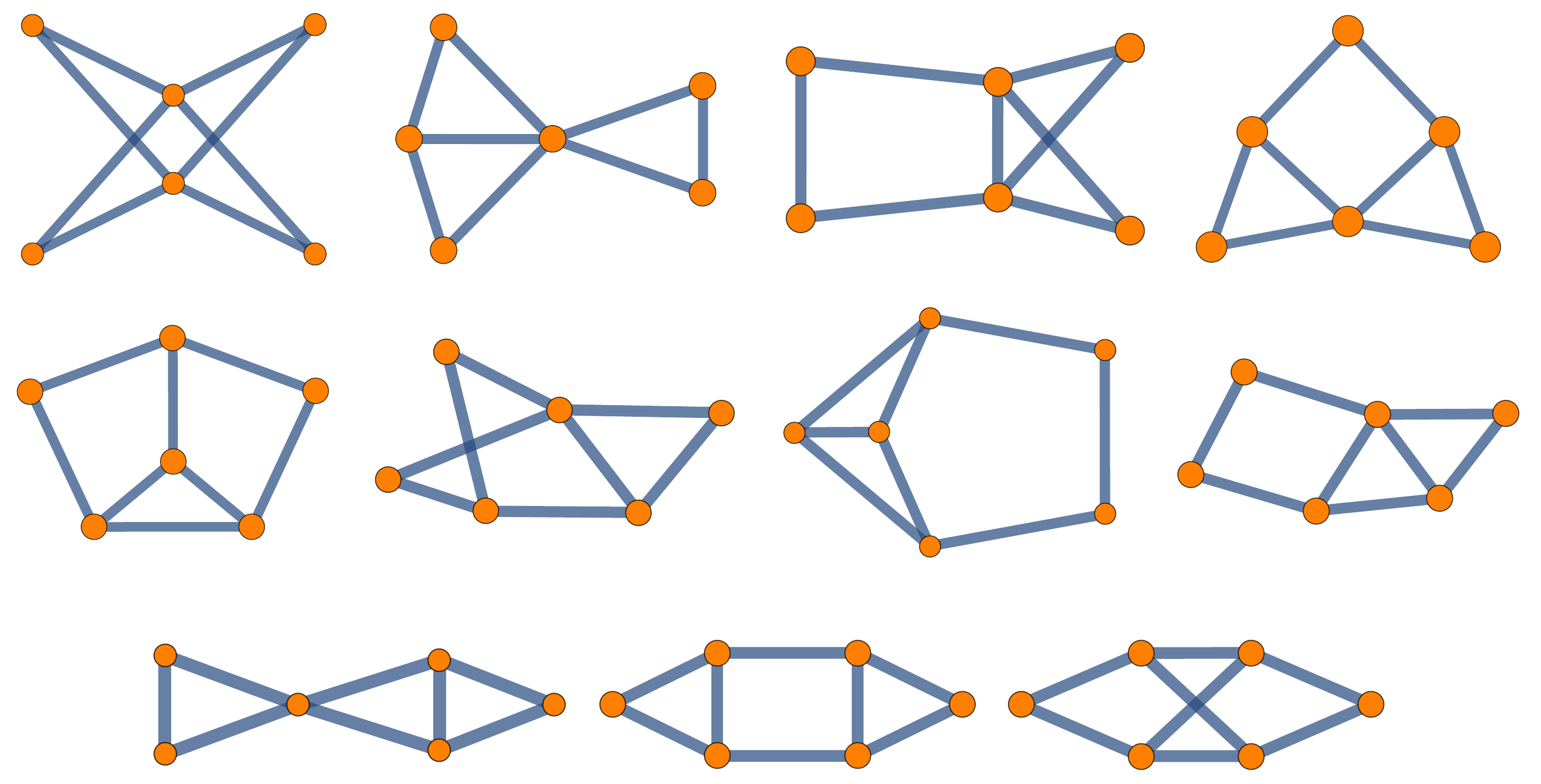}
\caption{The skeleton graphs at two loops generated by Sage. 
The final set of ribbon graphs is obtained from these by adding the colour factors and summing over all inequivalent assignments of external and internal points to the vertices.}
\label{fig:TwoLoopGraphs} 
\end{figure}

The first step in the  calculation of the four-point function is to
generate all relevant graphs. As discussed previously, at $\ell$ loops we need graphs with $4+\ell$
vertices and  $4+2\ell$ propagators. 
These graphs can be easily constructed at lower loops 
but the number of graphs increases very fast with
the loop order, so we used   
the program Sage \cite{Sage} to generate them.    
At one and two loops, the skeleton graphs are shown in figures  \ref{fig:OneLoopGraphs} and \ref{fig:TwoLoopGraphs} respectively, while  the number
of skeleton graphs up to four loops is shown in Table 
\ref{ta:NumberofGraphs}.    
\begin{table}[h!]
\centering
\begin{tabular}{ccccc}
loops & one & two & three & four   \\ 
\hline
\# of skeleton graphs & 3 & 11 & 63 & 513  
\end{tabular}
\caption{The number of skeleton graphs generated with Sage at each loop order. At four loops one can also draw two additional disconnected graphs, however they have to be discarded as they do not contribute to the connected correlator.} 
\label{ta:NumberofGraphs}
\end{table}
At one loop one  can draw graphs with either a single quartic vertex or with two trivalent vertices. The list of skeleton graphs can then be easily generated with Sage in both cases with the code
\begin{equation}
\begin{aligned}
&{\rm{graphs(5,degree\_sequence=[2,2,2,2,4])}} \,, \\
&{\rm{graphs(5,degree\_sequence=[2,2,2,3,3])}} \, . \\
\end{aligned}
\end{equation}   

The next step in the computation is to generate the ribbon graphs from the skeleton graphs obtained with Sage. In other words, each vertex of valence $v$ is assigned both a colour trace Tr$(T^{j_1} \ldots T^{j_v})$ and an $R^i_{j_1 j_2 \cdots j_v} $ factor as defined in (\ref{eq:defRij1jk}), and each propagator supplies an additional factor of $d_{ij}$. 
In general, each skeleton graph obtained with Sage leads to a large number of ribbon graphs. This happens because any inequivalent permutation 
of the indices $\{j_1,\ldots,j_v\}$ appearing in the colour traces and $R$ factors gives rise to a different ribbon graph. 
More precisely, a skeleton graph with $n$ vertices of valences $\{v_1,\ldots,v_n\}$ produces $\prod_{i=1}^n (v_i-1)!$ ribbon graphs, corresponding to the non-cyclic permutations at each vertex.

The graph propagators also provide colour delta functions so that the indices in the colour traces of the vertices are fully contracted. Effectively the color structure of the ribbon graph simplifies to a polynomial in $N_c$ through successive application of the fission and fusion rules
\begin{equation}
\begin{aligned}
{\rm{Tr}} \,  (T^a B T^a C) &= {\rm{Tr}} ( B) \, {\rm{Tr}} (C) 
- \frac{\gamma}{N_c} {\rm{Tr}}(BC) \, , \\
 {\rm{Tr}} (T^a B) {\rm{Tr}} (T^a C)  &=  {\rm{Tr}} (BC) 
- \frac{\gamma}{N_c} {\rm{Tr}}(B) {\rm{Tr}}(C) \, , 
\end{aligned} 
\end{equation}  
where $B$ and $C$ are arbitrary matrices and the cases $\gamma=0,1$ correspond to $U(N_c)$ and  $SU(N_c)$ respectively. We have checked that our
four-loop four-point correlator is independent of $\gamma$. This seems a bit surprising at the non-planar
loop level, but it is true for lenght-two operators.   
The $U(1)$ part of the gauge group is free and the 
external operators only give a trivial colour 
factor ${\rm{Tr}}( T^{a_1} T^{a_2}) = 
\delta^{a_1 a_2}$ in this case.\footnote{We thank Sergey Frolov for a discussion on this point.} For correlators involving 
half-BPS operators 
of higher weight, we expect that generically    
the result will depend on $\gamma$, see for example  Appendix A of \cite{HandlingII}.
The evaluation of ribbon graphs can be further 
simplified by using the property of the $R$ factor given in (\ref{eq:PermutationofTheIndices}), which allows us to rewrite all $R$ factors in a chosen canonical order, up to factors of $\langle \sigma_{ij} \sigma_{ik} \rangle$.
This implies that all ribbon graphs which originate from the same skeleton graph will produce the same canonical $R$ factors and differ only by a simple prefactor. In conclusion, to each skeleton graph we associate a canonical ribbon graph.  With this knowledge, and without performing any fermionic integration,
we could already rederive the well-known result 
that the correlator of four $\mathbf{20'}$ operators can only
receive a non-planar correction at four loops.
This happens because for lower loop orders all skeleton graphs produce the simple colour factor of
$(N_c^2-1)$. 

Finally, from the canonical ribbon graphs we  can
generate the final set of diagrams by associating vertices with external and internal positions in all inequivalent ways. Naively this 
would generate $(4+\ell)!$ graphs from each canonical
ribbon graph contributing to the $\ell$-loop correlator.  However, most graphs possess a non-trivial automorphism group, so that the number of inequivalent permutations is effectively much smaller.  
Moreover, once we have assigned positions to the vertices of the graph we can check if it is possible to produce an $(\theta^+)^4$ factor for each of the internal points, and it turns out that this test considerably decreases the number of allowed graphs. The number of final graphs for a generic polarization of the external operators is shown in the first line of table
\ref{ta:NumberofGraphsafteroperators}. Luckily, 
as explained in  subsection
\ref{sub:RulesTwistors}, we can perform the 
computation for
a particular choice of polarizations and reconstruct the 
final result unambiguously. 
If polarizations are such that only $y_{12}$ and $y_{34}$ are non-vanishing, then any graph with external-to-external propagators other than $d_{12}$ or $d_{34}$ is identically zero. This drastically reduces the number of diagrams to compute, as can be seen in table \ref{ta:NumberofGraphsafteroperators}.
    \begin{table}[h!]
\centering
\begin{tabular}{ccccc}
loops & one & two & three & four   \\ 
\hline
\# of graphs generic polarization & 45 &  1417 & 75141 &  6019618 \\
\hline
\# of graphs particular polarization & 1 & 73&  7939  & 
715350
\end{tabular}
\caption{The final number of diagrams which is obtained from canonical ribbon graphs by associating vertices to internal and external positions. In the generic polarization
all $y_{ij}$ are non-vanishing, while in the particular polarization any scalar product other than $y_{12}$ and $y_{34}$ is zero. The four-loop correlator in this paper was evaluated in the particular configuration, for which the number of graphs is greatly reduced.  }   
\label{ta:NumberofGraphsafteroperators}
\end{table}
 
After generating all the graphs and their prefactors, the final step in the calculation is to replace the $R$ factors by their expression \eqref{eq:defRij1jk}
and perform the fermionic integrations in $\theta^+$ at the Lagrangian insertions.  At four loops, all graphs
become a product of eight $R^{i}_{j_1 j_2 j_3}$ factors, thus producing
many terms with the correct number of $\theta_i^+$. 
An analytical computation is then very hard, especially because the dependence on the auxiliary twistor $\mathcal{Z}_{\diamond}$
only disappears after summing many graphs, see
\cite{Twistors} for details. 
Therefore, in this work we have performed the computation 
numerically  by giving integer values
to all components of the polarization and position vectors. While it is not necessary to restrict to integer numbers, we found that was a practical way of avoiding numerical fluctuations and errors.

The planar integrand had already been fixed in \cite{Integrand}, and we successfully reproduced their result up to four loops with our diagrammatic expansion in twistor space. This important comparison provides a good cross-check of our implementation.
At the non-planar level the integrand was written as a linear combination of four polynomials with undetermined 
coefficients. By computing each twistor diagram 
for twelve different sets of numerical values we produced an overcomplete system of equations and were able to fix those coefficients. Each numerical evaluation of the graphs took approximately 3 days 
in a single computer with 20 cores.  
The result is given in the following section, where we cross-check it further against some available
non-planar data.

If we consider configurations of points living in four dimensions, then there is a technical issue that arises. 
As discussed at the end of subsection \ref{sec:N4TwistorSpace}, the twistor 
action has the topological term 
$ i F \tilde{F}$ and so the evaluation of the graphs generates
terms involving the tensor 
$\epsilon^{\mu \nu \rho \lambda}$. These are spurious contributions at the level of the integrand and they must be absent in the final result, i.e. they necessary multiply functions that integrate to zero as $ i F \tilde{F}$ is a total derivative and it cannot
give any perturbative contribution.
Note that in Lorentzian signature the terms with an odd number
of epsilon tensors will be imaginary, which means that we can single out their contribution in our calculations. 
In order to fix them we found the complete basis of pseudoscalar conformal integrands with extended permutation symmetry. At first there seem to be many structures one can form, but when we implement permutation symmetry we must include minus signs to compensate the antisymmetry of the epsilon tensor, and at the end there are only 4 such polynomials. Furthermore, two of those correspond to pseudoscalar conformal Gram polynomials, leaving only two degrees of freedom.
 We obtained an imaginary non-planar component for five of our numerical data points, thus producing an overcomplete system of equations. Effectively, the result is a shift to the polynomial $P^{(1,4)}$ by     
\begin{align}\label{eq:polynomialcomepsilon} 
\tilde P^{(1,4)}(x_i) &= \frac{12 \,i}{14} \left(x_{17}^2 \,x_{18}^2 \,x_{26}^2 \,x_{28}^2 \,x_{35}^2 \,x_{37}^2 \,x_{45}^2 \,x_{4 6}^2 \,x_{78}^2 \,\epsilon^c_{123456} + S_8 \mbox{ perms}\right) \nonumber\\
&+ \frac{18 \,i}{14}( x_{18}^4 \, x_{27}^2 \,x_{28}^2 \, x_{36}^2 \,x_{37}^2 \, x_{45}^2 \,x_{47}^2 \,x_{56}^2\,\epsilon^c_{123456} + S_8 \mbox{ perms})\,.
\end{align}     
where we defined an $\epsilon^c$ tensor 
by using six-dimensional embedding vectors $(1,x^2,x^\mu)$       
\begin{equation}
\epsilon^c_{ijklmn}= \left|
\begin{array}{cccc}
	1 & 1 & \cdots & 1 \\
	x_i^2 & x_j^2 & \cdots & x_n^2\\
	x_i^\mu& x_j^\mu & \cdots & x_n^\mu
	\end{array}
	\right|\,.
\end{equation}     
By constructing the pseudoscalar polynomials with this six-dimensional epsilon tensor we guarantee that the polynomial in \eqref{eq:polynomialcomepsilon} is conformal.
In addition, 
recall that a product of ordinary epsilon tensors 
can always be reduced to a sum of products of Kronecker
deltas
\begin{equation}
\epsilon_{i_1 \ldots i_n} 
\epsilon^{j_1 \ldots j_n}
= n!\, \delta^{j_1}_{[i_1} \cdots \delta^{j_n}_{i_n]} \, .
\end{equation}
Thus, if spurious terms are generate in this way, 
the conformal and permutation symmetries ensure they can be written as a linear combination of the 32 four-loop polynomials
from \cite{Integrand}. 
We have to be confident that our numerical result is not polluted by this kind of spurious terms. 
As mentioned before, the terms generated with $\epsilon$ tensors have to integrate to zero, but we believe there is no such combination apart from the conformal Gram polynomials.
Thus we are confident that our twelve
data points still give an overcomplete set of
equations if we consider an extended basis which
have these additional polynomials that integrate to zero.             
Another check of our result can be made by noticing that the bosonic twistors, which are written in terms of $\lambda_{\alpha}$ and  
$x^{\dot{\alpha} \beta}$ due to the incidence relations (\ref{eq:incidencerelations}), can only 
contribute to the $R$-factors via the $\sigma^{\alpha}_{ij}$ defined in (\ref{eq:sigmadefinition}). 
Thus, all the matrices $\sigma^{\mu}_{\alpha \dot{\alpha}}$ are always contracted with a 
vector $x_i^{\mu}$. This implies that if a graph generates $\epsilon^{\mu \nu \rho \lambda}$ through simplifications
of products of $\sigma^{\mu}_{\alpha \dot{\alpha}}$, the resulting epsilon
tensor will always be fully contracted with 
a set of $x_i^{\mu}$'s and therefore 
vanish identically if all points are in a three-dimensional subspace. Thus these possible spurious terms belong         
to the space spanned by the 6 exclusively three-dimensional conformal Gram polynomials.
We have fit our numerical expressions to an extended basis consisting of the four non-planar polynomials defined in \eqref{eq:TheDefiningPolynomials} together with the 3d conformal Gram polynomials, but our result remained the same. 
Finally, we performed a few additional numerical tests
by checking  invariance under
a change of polarizations for the internal vertices
and independence of the result on 
the auxiliary twistor.

\section{OPE analysis}\label{sec:OPE}

The diagrammatic computation of the previous section fixed the non-planar polynomial to be
\begin{equation}
P^{(1,4)}(x_i) = -6\, Q_4 (x_i)\,,
\end{equation}
with $Q_4(x_i)$ defined in \eqref{Qs}. However, some of the terms in the polynomial lead to pseudo-conformal integrals. The  weight at each integrated point is $+4$, but those integrals are divergent in four dimensions and once we introduce dimensional regularization they lose their conformal properties. Since it is more cumbersome to deal with pseudo-conformal integrals, we chose to add a conformal Gram polynomial to the integrand and rewrite it as
\begin{equation}
P^{(1,4)}(x_i) = 2\sum_{j=1}^{32} \tilde q_j \,P_j^{(4)}(x_i)\,,
\end{equation}
with $P_j^{(4)}$ defined in equations (5.9) and (C.1) of \cite{Integrand} and the coefficients $\tilde q_j$ 
are given by
\begin{equation}\label{newq}
\tilde q= \{6, 6, -6, 8, 0, 6, 0, -1, -2, 0^2, 2, -1, 0^4, 2, 2, \
-2, -4, 0, -2, 0^3, -48, -4, 0, 4, 0, 0\}\,.
\end{equation}
The conformal Gram polynomials parametrize a three-dimensional subspace of the allowed polynomials, and we used two of its degrees of freedom to ensure that none of the pseudo-conformal integrals contributes in  \eqref{newq}. The last parameter is chosen in a way that eliminates some of the more difficult conformal integrals.

While the twistor space method allowed us to obtain the integrand of the correlation function, we are still faced with the integration of the Lagrangian insertions. Conformal symmetry reduces the complexity of the problem by restricting to functions of two cross-ratios only
\begin{equation}
u= \frac{x_{12}^2 x_{34}^2}{x_{13}^2 x_{24}^2} \,,\qquad\qquad v= \frac{x_{14}^2x_{23}^2}{x_{13}^2 x_{24}^2}\,.
\end{equation}
The four-loop ladder diagram is known exactly \cite{Ladder}, and recently the method of differential equations was used to fix a different topology \cite{DiffEden}, but in general the evaluation of four-loop four-point integrals is a daunting task.

Therefore, in this work we will focus on the Euclidean coincidence limit, which is tractable. We use the freedom of conformal symmetry to send $x_4$ to infinity, so that effectively we deal only with three-point integrals, and then we let $x_{1}$ approach $x_2$. In that case the cross-ratios become
\begin{equation}
u=\frac{x_{12}^2}{x_{13}^2}\rightarrow 0 \,, \qquad\qquad v=1-Y= \frac{x_{23}^2}{x_{13}^2}\rightarrow 1\,.
\end{equation}
Each of the conformal integrals has now two distinct scales $|x_{12}| \ll |x_{13}|$, which means that we can approximate the integrals with the method of asymptotic expansions. The idea is that for each integration variable $x_i$ we can now divide the integration domain into two regions, one where $x_{1i}$ is of the order of $x_{12}$, and another where it is of the order of $x_{13}$. In that way, an $\ell$-loop conformal integral will split into $2^\ell$ terms, corresponding to different distributions of the integration variables in the two regions. For example, if $x_{1i}$ is of the order of $x_{12}$ and $x_{1j}$ of the order of $x_{13}$, we can Taylor expand the propagator 
\begin{equation}\label{Expansion}
\frac{1}{x_{ij}^2} = \frac{1}{x_{1j}^2}\sum_{n=0}^\infty \left(\frac{2 x_{1i}\cdot x_{1j}-x_{1i}^2}{x_{1j}^2} \right)^n \,.
\end{equation}
This Taylor expansion is only convergent for  $|x_{1i}|\leq |x_{1j}|$, but we extend the integration domain in this region to the whole space. That makes the integrals strictly divergent, which can be resolved by introducing dimensional regularization and requiring that scaleless integrals vanish \cite{AsymptoticExpansion}. In order to find the lowest order in the small $u$ expansion we can ignore the $x_{1i}^2$ terms from the numerator of \eqref{Expansion}, and higher powers in the Taylor expansion will contribute to higher orders in the small $Y$ expansion. 

Looking at equation \eqref{Expansion} we realize that the two regions are effectively disentangled. This indicates that a term with a $\{k,\ell-k\}$ split of the integration variables into the  $\{x_{12},x_{13}\}$ regions will lead to a product of $k$- and $(\ell-k)$-loop two-point integrals, with external points $x_1$ and $x_2$, or $x_1$ and $x_3$, respectively.
This is a considerable simplification because we are able to approximate the conformal integral with single-scale integrals. The two-point integrals will generically have numerators with tensor structures, but we can follow the strategy of \cite{ThreeLoopEdenAsymptotic} to reduce them to scalar integrals. Once that is accomplished we use LiteRed \cite{LiteRed} and FIRE \cite{FIRE}  to implement IBP identities and reduce to a basis of simpler master integrals, which were obtained in \cite{Masters}. 

For simplicity we focused on a parametrization of the integrand which involved solely convergent integrals, but that was not strictly necessary. The method of asymptotic expansions can also be applied to pseudo-conformal integrals and the only caveat is that one must disregard conformal symmetry. More specifically, different terms in the integrand lead to integrals which are the same up to a permutation of the external points.  When they are convergent one can rely on the invariance of the cross-ratios under the group of double-transpositions
\begin{equation}
\{\mathrm{id}, (1 2)(3 4), (1 3)(2 4), (1 4)(2 3)\} 
\end{equation}
to relate them. However, the pseudo-integrals are divergent and in dimensional regularization they are not simply a function of cross ratios, which means that we must perform the asymptotic expansion independently for each permutation of the external points.

Since $\mathcal N=4$ SYM is a conformal theory, we can use the OPE to rewrite the four-point function as
\begin{equation}
G_4 = \frac{y_{12}^4 y_{34}^4}{x_{12}^4 x_{34}^4} \sum_{\Delta,l,n,m} C_{\mathcal O_{20'} \mathcal O_{20'}\mathcal O}^2 \;g_{\Delta,l}(u,v) \,Y_{n,m}(\sigma,\tau)\,,
\end{equation}
where $\sigma$ and $\tau$ are the $R$-symmetry cross-ratios, $Y_{n,m}$ is the $R$-symmetry block for the exchange of an operator in the $SU(4)$ representation $[n-m,2m,n-m]$, and $g_{\Delta,l}$ is the conformal block for an operator of dimension $\Delta$ and spin $l$. In the Euclidean OPE limit that was described above the conformal block simplifies to
\begin{equation}
g_{\Delta,l}(u,Y) \approx u^{\frac{\Delta-l}{2}} \,Y^l \,{}_2 F_1 \left(\frac{\Delta+S}{2},\frac{\Delta+S}{2};\Delta+S;Y\right)\,,
\end{equation}
which means that only the lowest-twist non-protected operators contribute to the leading $u$ behaviour of the four-loop correlator. Furthermore, a suitable choice of the external polarization vectors $y_i$ allows us to single out the $[0,2,0]$ representation. Since there is a single twist-two operator in the $20^{\prime}$ representation for each spin, we are able to extract all their OPE coefficients and anomalous dimensions.

The OPE data is  written as a double expansion on the genus and coupling constant
\begin{align}
\Delta(l) &= \Delta^{(0)}_l+ \sum_{g=0}^\infty \sum_{\ell=1}^\infty \frac{a^\ell}{N_c^{2g}}\gamma^{(g,\ell)}_l \,,\nonumber\\
C_{\mathcal O_{20'} \mathcal O_{20'}\mathcal O_{\Delta,l}}^2 &= \sum_{g=0}^\infty \sum_{\ell=0}^\infty \frac{a^\ell}{N_c^{2g}}\alpha^{(g,\ell)}_l \,.
\end{align}
Note that with the twistor space calculation we reconstruct the full $N_c$ dependence of the four-loop four-point function, which shows that the genus expansion truncates at the first non-planar order.  The non-planar anomalous dimensions are therefore
\begin{align}
\gamma_2^{(1,4)} &=-17280\, \zeta_5\,, \nonumber\\
\gamma_4^{(1,4)} &= 2800 + \frac{28000\, \zeta_3}{3} - \frac{100000\, \zeta_5}{3}\,,\nonumber\\
\gamma_6^{(1,4)} &= \frac{132986}{25} + \frac{85064 \,\zeta_3}{5} - \frac{230496 \,\zeta_5}{5}\,,\nonumber\\
\gamma_8^{(1,4)} &= \frac{220854227}{29400} + \frac{164142 \,\zeta_3}{7} - \frac{13898904\,\zeta_5}{245}\,.
\end{align}
These results match the perturbative computation of Velizhanin \cite{FourLoopNonPlanarKonishi,VelizhaninI,
	Velizhaninspin6} for spin 2, 4 and 6.\footnote{See footnote 1.}
%\footnote{There is in fact  a mismatch with the $\zeta_3$ component of the spin 6 anomalous dimension. Our result is 10 times that of Velizhanin, we hope this factor is simply due to a typo.}. 
Meanwhile the expression for spin 8 is new and its leading transcendental piece matches Velizhanin's conjecture for general spin
\begin{equation}
\gamma_l^{(1,4)} \Big|_{\zeta_5} = -7680 \,S_1(l)^2 \,,
\end{equation}
with $S_1(l)$ the harmonic sum. This seems to imply the behaviour  $\log^2 (l)$ for large spin, which is in contradiction wih the expected one. 
However, this is only part of the result and cancellations can occur at large spin.  
We also extracted the non-planar correction to the OPE coefficients, all of which are novel results,
\begin{align}
\alpha_2^{(1,4)} &=5760 \,\zeta_5 + 5040 \,\zeta_7 \,,\nonumber\\
\alpha_4^{(1,4)} &=\frac{112}{3} + \frac{400 \,\zeta_3}{9} + \frac{205040 \,\zeta_5}{441} + 600 \,\zeta_7 \,,\nonumber\\
\alpha_6^{(1,4)} &=\frac{48821149}{6534000} + \frac{41643 \,\zeta_3}{3025} + \frac{191044 \,\zeta_5}{9075} + \frac{
	588 \,\zeta_7}{11} \,,\nonumber\\
\alpha_8^{(1,4)} &=\frac{25811374441}{28171962000} + \frac{35869013 \,\zeta_3}{18632250} + \frac{
	477038734 \,\zeta_5}{2630252625} + \frac{3044 \,\zeta_7}{715}\,.
\end{align}
The transcendental structure of the spin 2 structure constant is quite interesting due to the absence of rational and $\zeta_3$ terms, but we do not currently understand why that happens. It is possible to derive 
a closed expression for the $\zeta_7$ part of the OPE coefficient, which is given by\footnote{The expression was found by  Gregory Korchemsky and we thank him
for communicating it to us.}   
\begin{equation}
\alpha_l^{(1,4)}\Big|_{\zeta_7} = 10080 \; \alpha_l^{(0,0)} \; S_1(l) \, ,
\end{equation}
with the tree-level planar coefficient given by $\alpha_l^{(0,0)} = 2 (l!)^2 / (2l)!$ .

Finally, we were able to perform the asymptotic expansion of the integrand up to spin 12, but the IBP reduction of the resulting two-point integrals was not possible. The data we have available is not sufficient to reconstruct an expression for generic spin in terms of harmonic sums, which means that we cannot access the large spin limit of the anomalous dimensions. Perhaps the method of differential equations can be used to find a higher-order expansion of the conformal integrals in a more efficient way.

\section{Conclusion}\label{sec:conclusion} 

In this work, we have fixed the four-loop non-planar  
integrand of the four-point function of
length-two half-BPS operators by a direct computation in twistor space. We performed the Grassmann integrations for numeric vales of the position and polarization vectors and fit the results obtained against a polynomial ansatz with four unknown coefficients. 
In this formalism each individual graph
preserves $\mathcal{N}=4$ superconformal symmetry,
apart from the reference twistor. The calculation 
was done with a particular choice of external polarizations in order to reduce the number of graphs, but the general result can
be unambiguously reconstructed due to the factorized dependence on the polarizations in equation 
(\ref{eq:DefinitionR1234}). In principle, 
one can also use our method to compute higher-point
and higher-loop integrands, both at the planar and non-planar level. However, the number of diagrams can grow a lot in those cases 
and the complexity of the fermionic integrations can also increase considerably. 
In our case, the integration took an average of three days on 20 cores for each set of numerical values.
In order to compute more complicated correlators it is perhaps necessary to implement a method which mixes numerical
and analytical methods. More specifically, one can first perform some of the fermionic
integrations analytically, as in the examples of 
Appendix \ref{sec:analytical}, and then complete the calculation numerically.

Let us now stress a technical detail of our method that could also show up for other
correlators. The twistor action from equation
(\ref{eq:twistoraction}) is chiral and contains the topological 
term $i F \tilde{F}$ when expanded in components. 
This implies that terms 
with $\epsilon^{\mu \nu \rho \lambda}$ can be generated
in twistor space calculations. 
In this work we were able to 
write down an ansatz for these terms and found
the contribution given in (\ref{eq:polynomialcomepsilon}).
This terms must integrate to zero and they do not
appear in the final integrand, but in principle it can be difficult to isolate the epsilon terms from numerical calculations. However,
when working in Lorentzian signature, the terms
with an odd number of epsilon tensors have an imaginary contribution and they
can be easily isolated. On the other hand, terms with an even number 
of epsilons give a real contribution which 
can be rewritten in the usual polynomial basis. 
Notice that these terms 
are identically zero when we restrict to a three-dimensial subspace, which implies that 
this possible spurious contribution can be written in terms of three-dimensional conformal Gram polynomials. 
A careful analysis using an extended basis showed that our result is not contaminated by this type of terms.       

In this work we focused on the non-planar 
corrections to the four-point function of $20^{\prime}$
operators, which start at four loops.  
There are also results in the literature for
correlation functions of length-$k$ half-BPS operators and for generic $k$ the non-trivial non-planar corrections usually start at lower 
loops.
At the planar level, all four-point functions of such operators are know 
up to five loops \cite{AllThreeLoop,
AllFiveLoop} while at the non-planar level
correlators of four length-$k$ operators are known 
up to two loops  only
\cite{Oneloopkkkk,TwoLoopkkkk}. 
It would be very interesting to compute 
non-planar corrections to more general four-point functions of
half-BPS operators at two and higher loops.
In that way we would further  test the integrability approach
to non-planar correlators developed in \cite{HandlingI,HandlingII},
and it could also help understand why non-planar corrections show up at specific loop orders in this approach. 
Maybe some of these computations can be done using 
the twistor reformulation of $\mathcal{N}=4$ SYM 
used in this paper  
together wih the prescription for composite operators in
twistor space of \cite{CompositeI,CompositeII,CompositeIII}. 

Finally,
it would be extremely nice to obtain a closed-form expression
for the non-planar anomalous dimension and 
structure constants of twist operators. This would allow to extract the non-planar cusp anomalous 
dimension and it would give valuable data for 
an integrability approach to the non-planar spectrum. 
We have made an OPE analysis of the four-point function 
and we obtained a few data points. However,
that data was not sufficient to completely fix the anomalous dimension at generic spin. It seems hard to push the OPE expansion further with the method of asymptotic expansions and so a new strategy is very likely needed. We hope to return to this point in the future.

\section*{Acknowledgement}

We thank Andrei Belitsky, Sergey Frolov, Paul Heslop and Gregory Korchemsky for useful discussions and 
especially acknowledge
Vasco Gon\c{c}alves for suggesting the project and for participating at an early stage. 
We also thank Vasco Gon\c{c}alves and  
Gregory Korchemsky for valuable comments on the manuscript. 
T.F would like to thank the warm hospitality of the Arizona State
University where this work was finished.  
This work was supported by the Serrapilheira 
Institute (grant number Serra-1812-26900).  R.P. is supported by SFI grant 15/CDA/3472.

\appendix
%%%%%%%%%%%%%%%%%%%%%%%%%%%

%\section{Multi-Particle Corrections at Two Loops}

%%%%%%%%%%%%%
 \section{Conventions}\label{sec:Conventions}

 In this work, we have used the following conventions 
 to raise and lower $SU(2)$ indices  
 \begin{equation}
x_{\alpha\dot{\alpha} } = x_{\mu} \sigma^{\mu}_{\alpha\dot{\alpha} } \, , \quad  \quad x^{\dot{\alpha} \alpha} = \epu{\alpha\beta} x_{\beta \dot \beta} \epu{\dot \beta \dot \alpha} \, , \quad \quad 
x_{\alpha\dot \alpha} =\epl{\dot \alpha \dot \beta} x^{\dot \beta \beta}  \epl{\beta\alpha} \, . 
\end{equation}
Similarly, for the $y^{\mathfrak{a}}_{\mathfrak{b^{\prime}}}$  variables introduced in (\ref{eq:DefinitionyTwoIndices}) as a parametrization of the harmonic variables we have
\begin{equation}\label{ys}
y^{\mathfrak{a^{\prime}}}_{\mathfrak{b}} = 
y^{\mathfrak{a}}_{\mathfrak{b^{\prime}}} \epsilon^{\mathfrak{b^{\prime} \mathfrak{a^{\prime}}}} 
\epsilon_{\mathfrak{a} \mathfrak{b}} \, , \quad \quad 
y^2 = - y^{\mathfrak{b}}_{\mathfrak{a^{\prime}}} 
y^{\mathfrak{a^{\prime}}}_{\mathfrak{b}}/2 \, .  
 \end{equation} 
The epsilon tensors are defined with $\epl{12}=\epu{12}=1$, so that they obey  
\begin{equation}
\epl{ab} \epu{a c} = \delta_b^c \, .
\end{equation}
For both the spacetime and $R$-symmetry matrices we use the following short-hand notation
\begin{equation}
x^{ \dot{\alpha} \alpha}_{ij} = x^{ \dot{\alpha} \alpha}_i - x^{ \dot{\alpha} \alpha}_j \, ,  \quad {\rm{and}} \quad 
(y_{ij})^{\mathfrak{a}}_{\mathfrak{b^\prime}} =
(y_{i})^{\mathfrak{a}}_{\mathfrak{b^\prime}}-
(y_{j})^{\mathfrak{a}}_{\mathfrak{b^\prime}} \, . 
\end{equation}  

Using the properties of the Pauli matrices it is possible to show that 
%\begin{equation}
%(\s^\mu)^{\dot \alpha \alpha} \s^\nu_{\alpha\dot %\alpha} = -2 \eta^{\mu\nu}
%\end{equation}
\begin{equation}
x^{\dot \alpha}_\alpha \,y^\alpha_{\dot{\alpha}}= -2 \,x\cdot y \, , \quad \quad x^{\dot \alpha \alpha} x^{\dot \beta}_\alpha = x^2 \epu{\dot \alpha \dot \beta} \, , 
\end{equation}
%Finally, since $x^{\dot \alpha \alpha} x^{\dot \beta}_\alpha$ is antisymmetric in $\dot \alpha$ and $\dot \beta$, then it must be proportional to $\epu{\dot{\alpha}\dot \beta}$, from which we can show that 
%\begin{equation}
%x^{\dot \alpha \alpha} x^{\dot \beta}_\alpha = x^2 \epu{\dot \alpha \dot \beta}
%\end{equation}
We can manipulate the last equation above to see that 
\begin{equation}
x^{\dot\alpha \alpha} \left(\frac{1}{x^2} x_{\alpha\dot\gamma}\right)= \delta^{\dot{\alpha}}_{\dot{\gamma}} \, , \quad 
\rightarrow \, \quad 
(x^{-1})_{\alpha\dot{\alpha}}= \frac{1}{x^2} x_{\alpha\dot \alpha} \, .
\end{equation}
Analogously, one has
\begin{equation}
(y^{-1}_{ij})^{\mathfrak{a}^{\prime}}_{\mathfrak{b}}
= \frac{1}{y_{ij}^2} (y_{ij})^{\mathfrak{a}^{\prime}}_{\mathfrak{b}} \, . 
\end{equation}

Concerning 
the integrations of the Grassmann variables $\theta_{i,\alpha}^{\pm a}$,
we use the convention
\begin{equation}
\int \mathrm d^4 \theta_i^\pm=\int \mathrm d \theta^{\pm 1}_{i,1} \mathrm d \theta^{\pm 2}_{i,1} \mathrm d \theta^{\pm 1}_{i,2} \mathrm d \theta^{\pm 2}_{i,2} \, . 
\end{equation}
This implies that  $\theta^{\pm 1}_1 \theta^{\pm 2}_1 \theta^{\pm 1}_2  \theta^{\pm 2}_2$ integrates to one.
 
\section{Examples of fermionic integrations} 
\label{sec:analytical}

In this work we have computed all  necessary diagrams numerically, i.e we have set all components of the position and polarization vectors to integer values.
Nevertheless, for a given diagram it might be possible to perform some or even all the fermionic integrations analytically.
Eventually one can combine the two methods 
in an efficient way and reduce the complexity of many graphs.  
Here, we give some examples of integrations that can be performed easily.
First, consider the $\theta_i^+$ integration when the internal point $i$ is a bivalent vertex. In that case the Grassmann variables can be found in the product of $R$ factors from the adjacent vertices $j$ and $k$
\begin{equation}
 R^j_{a_1 \ldots a_m i } R^k_{b_1 \ldots b_n i}
\, .
\end{equation}
From equation (\ref{eq:defRij1jk}) we can see that the relevant terms for the integration of $\theta_i^+$ come solely from
\begin{equation}
\delta^2(\brst{ki}{k}{-}+  A_{ki}) \delta^2(\brst{ji}{j}{-}+ A_{ji}) \, ,
\end{equation}
which means that the only term with four $\theta_i^+$ is
\begin{equation}
\brst{ik}{i}{+\mathfrak{a}} \yinv{ki}^{ 1^{\prime}}_{\mathfrak{a}} \,  \brst{ik}{i}{+\mathfrak{b}} \yinv{ki}^{2^{\prime}}_{\mathfrak{b}} \,
\brst{ij}{i}{+\mathfrak{c}} \yinv{ji}^{1^\prime}_{\mathfrak{c}} \,  \brst{ij}{i}{+\mathfrak{d}} \yinv{ji}^{ 2^\prime}_{\mathfrak{d}}
= (\theta_i^+)^4\, \frac{\brss{ik}{ij}^2}{y_{ki}^2 y_{ji}^2} \, . 
\end{equation}
Therefore, the fermionic integration gives
\begin{equation}
\int \mathrm d^4 \theta_i^+ R^j_{a_1 \ldots a_m i } R^k_{b_1 \ldots b_n i} = R^j_{a_1 \ldots a_m} R^k_{b_1 \ldots b_n} \frac{\brss{ik}{ij}^2 }{y_{ki}^2 y_{ji}^2} \frac{\brss{j a_m}{j a_1}}{ \brss{j a_m}{j i} \brss{ji}{ja_1}} \frac{\brss{k b_n}{k b_1}}{ \brss{k b_n}{k i} \brss{ki}{kb_1}} \, .
\end{equation}
%Crucially, the corresponding graph also comes with %the propagator factor $d_{ij} d_{ik}$,  which contains %$y_{ij}^2 y_{ik}^2$ and cancels the denominator above. 

Another example of a simple $\theta_i^+$ integration consists of the internal point sitting at a trivalent vertex connected to two bivalent vertices $j$ and $l$ and one higher-valence denoted by $k$. In that case the necessary fermionic variables are provided by the product of two $R$ factors which integrates to
\begin{equation}
\int d^4 \theta_i^+ \, R^i_{jkl} \,R^k_{a_1 \ldots a_n i} =R^k_{a_1\ldots a_n} \frac{y_{jl}^2 \brss{ij}{ik}\brss{ik}{il}}{y_{ij}^2 y_{ik}^2 y_{il}^2 \brss{ij}{il}} \frac{\brss{ka_n}{ka_1}}{\brss{ka_n}{ki}\brss{ki}{ka_1}} \,.
\end{equation} 

As the final example, let us now consider the $\theta_i^+$ integration when the internal point $i$ is a quartic vertex connected to bivalent vertices only.
In that case the Grassmann variables originate from the $R$ factor at the internal vertex $i$ and we have 
\begin{equation}\label{eq:analiticalBotie}
	\int \mathrm d^4 \theta_i^+ \,R^i_{1234} =\frac{1}{y_{i1}^2 y_{i2}^2 y_{i3}^2  y_{i4}^2 } \left(y_{12}^2 y_{34}^2 \frac{\brss{i1}{i3}  \brss{i2}{i4} }{\brss{i1}{i2}  \brss{i3}{i4} }+y_{13}^2 y_{24}^2  +y_{14}^2 y_{23}^2 \frac{  \brss{i1}{i3}\brss{i2}{i4}  }{ \brss{i1}{i4} \brss{i2}{i3}  } \right) \,.
\end{equation}

\end{document}